\documentclass[twocolumn,prb,showpacs,preprintnumbers,amsmath,amssymb,floatfix,superscriptaddress]{revtex4}
\usepackage{graphicx}
\usepackage{amsmath}
\usepackage{amssymb}
\usepackage{color}

\begin{document}

\title{Transport properties of nonhomogeneous segregated composites}

\author{B. Nigro}\email{biagio.nigro@epfl.ch}\affiliation{LPM, Ecole Polytechnique F\'ed\'erale de Lausanne, 
Station 17, CP-1015 Lausanne, Switzerland}
\author{G. Ambrosetti}\affiliation{LPM, Ecole Polytechnique F\'ed\'erale de
Lausanne, Station 17, CP-1015 Lausanne, Switzerland}
\author{C. Grimaldi}\email{claudio.grimaldi@epfl.ch}\affiliation{LPM, Ecole Polytechnique F\'ed\'erale de
Lausanne, Station 17, CP-1015 Lausanne, Switzerland}
\author{T. Maeder}\affiliation{LPM, Ecole Polytechnique F\'ed\'erale de
Lausanne, Station 17, CP-1015 Lausanne, Switzerland}
\author{P. Ryser}\affiliation{LPM, Ecole Polytechnique F\'ed\'erale de
Lausanne, Station 17, CP-1015 Lausanne, Switzerland}
%\date{31-08-2009}

%\widetext
\begin{abstract}
In conductor-insulator composites in which the conducting particles are dispersed in an insulating
continuous matrix the electrical connectedness is established by interparticle quantum tunneling.
A recent formulation of the transport problem in this kind of composites treats each conducting particle
as electrically connected to all others via tunneling conductances to form a global tunneling network. 
Here, we extend this approach to nonhomogeneous composites with a segregated distribution of the 
conducting phase. We consider
a model of segregation in which large random insulating spherical inclusions forbid small
conducting particles to occupy homogeneously the volume of the composite, and allow tunneling
between all pairs of the conducting objects. By solving numerically the corresponding tunneling
resistor network, we show that the composite conductivity $\sigma$ is enhanced by segregation and
that it may remain relatively large also for very small values of the conducting filler concentration.
We interpret this behavior by a segregation-induced reduction of the interparticle distances, which is
confirmed by a critical path approximation applied to the segregated network. Furthermore, we identify
an approximate but accurate scaling relation permitting to express the conductivity of a segregated systems
in terms of the interparticle distances of a corresponding homogeneous system, and which provides an explicit formula
for $\sigma$ which we apply to experimental data on segregated RuO$_2$-cermet composites.
\end{abstract}
\pacs{64.60.ah, 73.40.Gk, 72.80.Tm,  72.20.Fr}
\maketitle

\section{Introduction}
\label{intro}

\begin{figure*}[t]
\begin{center}
\includegraphics[scale=1.4,clip=true]{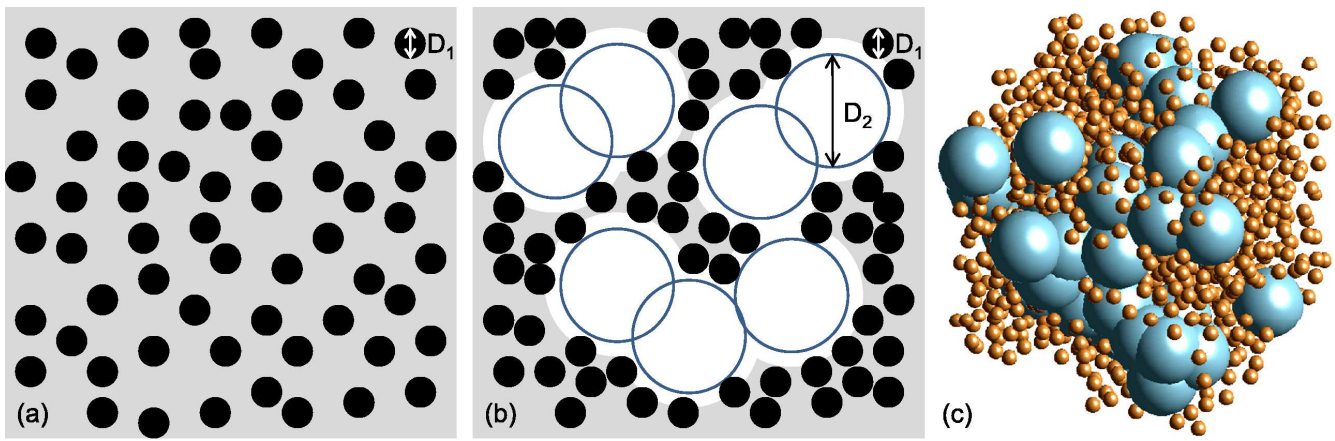}
\caption{(Color online) Two-dimensional representation of (a) an homogeneous and (b) segregated dispersion
of the conducting particles in the continuum.
The insulating (conducting) inclusions are represented by open (filled) circles of diameter $D_2$ ($D_1$).
The insulating fillers can penetrate each other, while the conducting particles are impenetrable with
respect to themselves and to the insulating ones. The grey region represents the space available for
placing the centers of the conducting spheres. (b) Example of a distribution of insulating and
conducting particles with $D_2/D_1=4$.}\label{fig1}
\end{center}
\end{figure*}

The transport properties of two-phase heterogeneous materials are strongly related to the structure of the composite i.e. the volume fraction $\phi$ of the conductive fillers, their size and shape, and their dispersion into the
insulating medium.\cite{Nan2010,Schilling2010}
Controlling the conductivity $\sigma$ of disordered composite materials by tuning these structural parameters
is of fundamental importance for several applications in the fields of micro and nanoelectronics such as,
to name a few, electromagnetic interference shielders, resectable fuses, strain and chemical sensors,
flexible conductors, and anti-static compounds.\cite{devices}

In several classes of composite materials in which the conductive particles are dispersed
into a insulating medium, like e.g. polymer-based composites or metal-glass cermets, the prominent
transport mechanism is quantum tunneling. In this case of tunneling, the conductance between two particles decays exponentially with the inter-particle distance over a characteristic tunneling length $\xi$ which is of the order of a fraction to a few nanometers depending on the material characteristics.

Typically, the tunneling mechanism is approximated by treating the fillers as core-shell objects, where the impenetrable hard core represents the physical particle and the thickness of the concentric penetrable shell is identified
with $\sim \xi$.\cite{Wang1993,Jing2000,He2004,Berhan2007} In this way, the overall behavior of the composite conductivity $\sigma$ as a function of
$\phi$ is commonly interpreted in the framework of percolation theory,\cite{Stauffer1994,Sahimi2003} which considers two given conducting particles as either electrically connected or disconnected if their mutual distance is respectively
lower or larger than the cut-off length imposed by the shell thickness.
Under this assumption, the system undergoes a conductor-insulator transition at a critical concentration $\phi_c$ of conducting phase, which is associated with the formation of a globally connected cluster of electrically linked filler particles which spans the entire sample. Below $\phi_c$ there is no such sample-spanning cluster and
$\sigma$ is zero, while for $\phi>\phi_c$ the conductivity follows a power law behavior of the form
$\sigma \propto (\phi-\phi_c)^t $, where $t$ is a critical exponent.

Despite of its simplicity and of the unquestionable insights that it provides on the transport 
problem,\cite{Balberg1984,Schilling2007,Kyrylyuk2008,Otten2009} 
introducing the concept of a cut-off through which electrical connections are established, is nevertheless a too crude 
approximation which alters the real nature of the interparticle electrical
connectedness. This is indeed characterized by the fact that, due to the tunneling mechanism, two given
particles are always electrically connected regardless of their mutual distance,\cite{Balberg2009} even if
the strength of the connection (i.e., the interparticle conductance) decays exponentially with such distance.

Recently, we have reformulated the transport problem in conductor-insulator composites by allowing each
conducting particle to be connected to all others via tunneling processes,\cite{Ambrosetti2009,Ambrosetti2010a,Ambrosetti2010b} and so without imposing the restrictive hypothesis on which the usual core-shell model is based.
By explicitly taking into
account the composite morphology or microstructure and the conducting particles shape and dimensions,
this global tunneling network (GTN) model is able to describe the overall conductivity dependence upon the
particle concentration for many different classes of composites, ranging from granular-like systems to colloidal
dispersions in a continuum insulating matrix.\cite{Ambrosetti2010b}
When applied to the many published data on colloidal nanotube, nanofiber, nanosheets, and nanosphere composites, this formalism has permitted to extract important
microscopic properties, such as the tunneling decay length $\xi$, directly from the experimental
conductivity vs concentration curves.\cite{Ambrosetti2009,Ambrosetti2010a}

In this paper we apply the GTN model to composites where the dispersion of the conducting fillers in the
insulating continuum is not homogeneous. Specifically, we consider systems
in which the insulating phase forbids the conducting fillers to occupy large (compared to the particle size)
volumes inside the material, thereby leading to a segregated spatial distribution
of the conducting phase.\cite{He2004,Kusy1977,Johner2009}
In real composites, like e.g. RuO$_2$-based cermets,\cite{RuO2}
this is achieved when the size of the insulating grains is consistently larger then the one of the
conducting fillers, and thermal treatments, inducing softening and sintering of the insulating phase
without large-scale mixing, lead to a segregated distribution of the conducting particles in a continuous
insulating matrix. Besides the already cited RuO$_2$-based cermets, another important class of segregated 
conductive composites is that of polymer-based ones,\cite{Malliaris1971,Mamunya2002} where an inorganic or 
carbonaceous conductive filler is mixed with significantly larger polymer particles and the resulting 
compound is molded.

The study of the transport properties in segregated composites is important for many technological
applications where low filler concentrations are demanded in order to have high conduction regimes
combined with the unaltered mechanical properties of the host insulating medium, 
or to reduce the quantity of the conductive phase when its cost is high.
So far, the problem
of conduction in segregated systems has been limited to the evaluation of the critical concentration
$\phi_c$ within the percolation framework, and both
lattice\cite{Kusy1977,Malliaris1971,lattices} and continuum\cite{He2004,Johner2009} models
have evidenced that $\phi_c$ is (usually, see Ref.\onlinecite{Johner2009}) lowered by segregation.
In the following we shall go beyond the
percolation approach by considering the GTN scenario for the segregation problem and by
solving numerically the tunneling network equations. We find that the composite
conductivity $\sigma$ for fixed filler concentration can be strongly enhanced by the segregation,
in accord with the observed trends. Furthermore, we show that the calculated filler dependencies of $\sigma$
can be reproduced to a great accuracy by the critical path approximation,\cite{Ambegaokar1971} which we find to follow
a simple scaling law permitting us to provide analytical formulas for $\sigma(\phi)$. When applied to
experimental data of real segregated composites, our formulas can be used to estimate the degree of segregation
of the composite and the value of the tunneling length $\xi$.

The structure of the paper is as follow: in Sec.~\ref{model} we present our GTN model for segregated
composites and in Sec.~\ref{conduct} we calculate numerically the composite conductivities.
In Sec.~\ref{critical} we present our results on the critical tunneling distance which will be used to
approximate the numerical results of Sec.~\ref{conduct} and to provide explicit formulas
for the conductivity. These are applied in Sec.~\ref{applic} to some previously published
data of segregated composites to extract the tunneling distance.
Section \ref{concl} is left to the conclusions.

\section{Model}
\label{model}

We model the conductor-insulator composite as described in Fig.~\ref{fig1}(a) and (b) for the case of an homogeneous dispersions of conducting particles in the continuum and for a segregated distribution, respectively.
In Fig.~\ref{fig1}(b) the spherical
particles of diameter $D_2$ represent the insulating inclusions (e.g., the glassy frit particles in RuO$_2$ cermets)
while the conducting particles are modeled as hard spheres of diameter $D_1$. The two kinds of particles
are mutually impenetrable and, furthermore, we assume that the $D_2$ spheres can penetrate each other in order to simulate for instance sintering and softening of the insulating grains.
Typically, as in RuO$_2$ cermets, $D_2$ is as large
as a few micrometers while $D_1$ ranges from tens to hundreds of nanometers, so that the regime $D_2\gg D_1$ is
the one of practical interest. Keeping this in mind, we shall consider in the following also moderate values of $D_2/D_1$ to better appreciate the overall trends towards the $D_2/D_1\gg 1$ regime.

The system is generated by first randomly placing the penetrable insulating spheres into a three dimensional
cubic volume of side length $L$ with a given number density $\rho_2 = N_2/L^3$, where  $N_2$ is the number of
$D_2$ particles and $L$ is chosen to be at least one order of magnitude larger than $D_2$ (we assume
periodic boundary conditions). Since the positions
of the insulating spheres are uncorrelated, their fractional volume is $\phi_2 = 1-\exp(-v_2\rho_2)$
where $v_2 = \pi D_2^3/6$ is the volume of a single sphere.\cite{Torquato2002}
After having placed the insulating spheres, $N_1$ conducting hard spheres are added to the system
through random sequential addition (RSA), where random placing is accepted only if there is no
overlap with the other $D_1$ and $D_2$ spheres.
The RSA procedure is repeated until the desired volume fraction value $\phi_1=\rho_1 v_1$,
where $v_1=\pi D_1^3/6$ and $\rho_1 = N_1/L^3$, is reached.
However, since the conducting spheres cannot penetrate the
insulating ones, the available volume for placing the centers of the $D_1$ particles is reduced by
the factor\cite{Johner2009}
\begin{equation}
\label{eq:phiav}
\upsilon^* = (1-\phi_2)^{(1+D_1/D_2)^3}
\end{equation}
with respect to the total volume $L^3$ of the system. This defines an effective volume
fraction $\phi_1/\upsilon^*$, larger than $\phi_1$, for the conducting spheres.
Hence, for small values of $\upsilon^*$, the limit achievable through
RSA can become much lower than that of the homogeneous limit at $\upsilon^*=1$,
which is $\phi_1^{\rm max}\simeq 0.382$,\cite{Sherwood1997} and scales approximately as
$\phi_1^{\rm max}\upsilon^*$. For densities larger than the RSA limit we
have considered cubically arranged initial configurations where particles
overlapping with the insulating spheres were removed.
Both the initial RSA and the cubic configurations were then relaxed via Monte Carlo runs, where random
displacements of the $D_1$ sphere centers were attempted and accepted only if they did not
overlap with any of its neighbors and with the $D_2$ particles. Equilibrium was considered attained
when the mean nearest-neighbor distances between the $D_1$ particles did not change within
statistical errors upon further Monte Carlo displacements. An example of the so-obtained distribution
of $D_1$ and $D_2$ spheres is shown in Fig.~\ref{fig1}(b) for the case $D_2/D_1=4$.

In describing the overall conductivity arising from the system described above
we go beyond the usual core-shell approximation, and employ the GTN model to the subset
of $D_1$ particles of the composite. Hence, we treat any two conducting spheres
centered at $\mathbf{r}_i$ and $\mathbf{r}_j$ as electrically connected through tunneling
processes, irrespectively of their mutual distance $r_{ij} = | \mathbf{r}_i - \mathbf{r}_j |$.
By assuming that the particle size $D_1$ and the temperature are large enough to neglect charging
energy effects, then the inter-particle conductance is given by:
\begin{equation}
\label{eq:tunnel}
g_{ij} = g_0 \exp\!\left(-\frac{2 \delta_{ij}}{\xi}\right)
\end{equation}
where $g_0$ is a constant ``contact'' conductance which we shall set equal to the unity,
$\xi$ is the tunneling decay length, and $\delta_{ij}=r_{ij}-D_1$
is the minimal distance between the surfaces of two  conducting spheres.
For a system composed by $N_1$ particles, the GTN model is then equivalent to a weighted random network with
$N_1$ nodes, each with coordination number $N_1-1$. However, contrary to the usual models of weighted
networks,\cite{Li2007} the weight of each link is not random but it is given by Eq.~\eqref{eq:tunnel}, which
depends on the particular arrangement of the conducting fillers in the composite. This characteristics
of the model permits in principle to study on equal footing composites with different statistical properties
of the microstructure, and has been successfully applied to homogeneous colloidal and granular
composites.\cite{Ambrosetti2009, Ambrosetti2010a,Ambrosetti2010b}
In the following we shall show that the GTN approach is also able to describe the conductivity of segregated
systems, thus providing a theoretical framework for the study of inhomogeneous composites.

\section{Conductivity}
\label{conduct}

\begin{figure}[t!]
\begin{center}
\includegraphics[scale=0.36,clip=true]{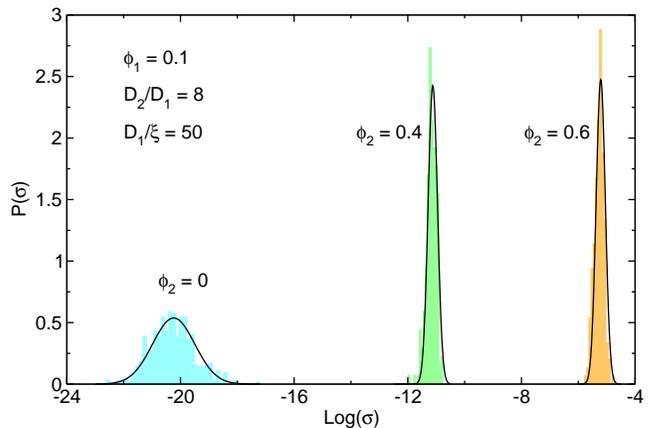}
\caption{(Color online) Histograms for the distribution $P(\sigma$) of the conductivity obtained
from $500$ realizations of the tunneling network for $D_2/D_1=8$, $D_1/\xi=50$ and $\phi_1=0.1$.
The volume fraction values of the insulating spheres are $\phi_2=0$ (with $L/D_1=25$)
and $\phi_2=0.4$, $0.6$ (with $L/D_1=80$).
The solid lines are fits to log-normal distribution functions.}\label{fig2}
\end{center}
\end{figure}

\begin{figure*}[t!]
\begin{center}
\includegraphics[scale=0.7,clip=true]{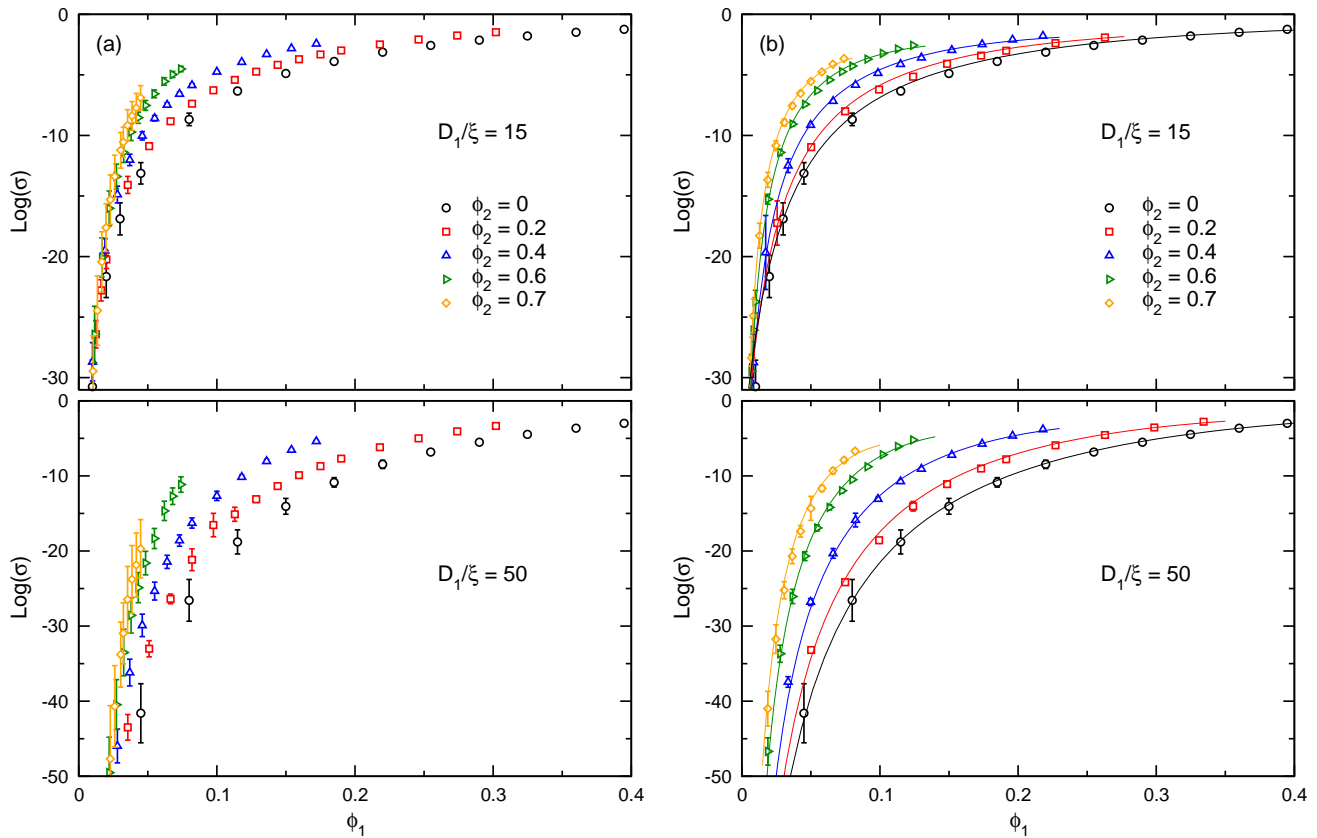}
\caption{(Color online) Calculated GTN conductivity as a function of the volume fraction $\phi_1$ of the
conducting spheres with diameter $D_1$ for $D_1/\xi=15$ and $50$, and for different values of the volume
fraction $\phi_2$ of the insulating spheres with diameter $D_2$. (a) $D_2/D_1=4$, (b) $D_2/D_1=8$.
The solid lines in (b) are fits of our formula.}\label{fig3}
\end{center}
\end{figure*}

In this section we present the results of our numerical calculations of the filler concentration dependence of
the conductivity $\sigma$ for different segregated systems specified by $\phi_2$ and $D_2/D_1$.
Since in the GTN model each conducting particle is connected to all others through Eq.~\eqref{eq:tunnel},
the calculation of $\sigma$ would require the solution of a network with $N_1(N_1-1)/2$ resistors, which
is a computationally demanding, or even insurmountable, task for the $N_1$ values considered in our study.
However, we can exploit the exponential decay of Eq.~\eqref{eq:tunnel} by neglecting
contributions from tunneling between particles sufficiently far away apart. Indeed, depending on the value
of $\xi$ and of the filler concentration $\phi_1$, it is possible to identify an upper artificial cut-off
$\delta^{*}$ such that the conductances between particles at mutual distances $\delta_{ij} > \delta^{*}$
can be safely removed, reducing drastically the number of connected particles in the network.
In our calculations we have then chosen $\delta^*$ such that, depending on $\phi_1$, $\exp(-2\delta^*/\xi)$
is from five to twenty orders of magnitude smaller than the overall network conductivity.\cite{deltastar}

Once the network has been reduced, we have evaluated $\sigma$ by combining the numerical decimation algorithm
of Ref.~\onlinecite{Fogelholm1980} with a preconditioned conjugated gradient method.
Specifically, we decimated iteratively the network starting from the nodes with the lowest coordination number
in order to eliminate dead ends and to compact the network. We continued the decimation procedure until a single
conductance was left, whose value coincided with the conductance of the original network.
If the computational time for the node decimation was too large, as it was typically the case for segregated systems
with $\phi_1/\upsilon^*$ large, we switched to the conjugate gradient method (see e.g. Ref.~\onlinecite{Batrouni1988})
applied to the partially decimated network. 
We have applied this procedure to $N_r=200-600$ realizations of systems
with $N_1$ conducting spheres ranging from $N_1=250$ (for $L/D_1=50$ and $\phi_1=10^{-3}$) to $N_1=322690$
(for $L/D_1=80$ and $\phi_1=0.33$).

In Fig.~\ref{fig2} we show the distributions $P(\sigma)$ of the $\sigma$ values obtained from $500$ realizations
of the system with $\phi_1=0.1$,
$D_1/\xi=50$, $D_2/D_1=8$, and for $\phi_2=0$ (with $L/D_1=25$), $0.4$, and $0.6$ (with $L/D_1=80$).
All three sets of data follow approximately a log-normal distribution (solid lines) which stems from the
exponential decay of Eq.~\eqref{eq:tunnel} (note that the distribution of the $\phi_2=0$ case is
broader than the two others because of the smaller size of $L$). The distributions are peaked at the
average of the logarithm of $\sigma$, with no significant drifts when the system size is increased for fixed $\phi_1$.
It is clear from the figure that, although $\phi_1$ is kept fixed, the mean value of the conductivity
steadily increases as the volume fraction $\phi_2$ of the insulating spheres is enhanced.
From Eq.~\eqref{eq:phiav} this trend can be interpreted by noticing that as $\phi_2$ increases, the available
volume fraction $\upsilon^*$ decreases, so that the conducting particles occupy a reduced volume compared to
the homogeneous $\phi_2=0$ case (i.e., the effective concentration $\phi_1/\upsilon^*$ is larger).
In turns this means that, as it can be inferred from Figs.~\ref{fig1}(a) and (b), the mean interparticle
distances $\delta_{ij}$ in Eq.~\eqref{eq:tunnel} are
reduced for $\phi_2\neq 0$, thereby leading to an enhancement of the overall conductivity.

This behavior is clearly shown in Fig.~\ref{fig3}, where we plot $\sigma$ (symbols) as a function of
$\phi_1$ for $D_2/D_1=4$, Fig.~\ref{fig3}(a), and $D_2/D_1=8$, Fig.~\ref{fig3}(b), and for several
values of the insulating sphere densities $\phi_2$.
The reduction in $\sigma$ for decreasing $\phi_1$, which is a direct consequence of the fact that as $\phi_1$
is reduced the interparticle distances $\delta_{ij}$ get larger, can be strongly mitigated
by the segregation which, through the reduction of $\upsilon^*$, tends instead to decrease $\delta_{ij}$.
We have therefore that, as shown in Fig.~\ref{fig3}, for fixed values of the tunneling length $\xi$, as the segregation
is enhanced the threshold value of $\phi_1$ required to achieve a given $\sigma$ decreases considerably.
By combining this result with the observation that, in practice, the lowest measurable conductivity
$\sigma_{\rm min}$ in real composites is set either by the experimental setup or by the intrinsic conductivity
of the insulating phase, and that this defines a sort of ``critical" threshold $\phi_{1c}$ at which
$\sigma\sim\sigma_{\rm min}$,\cite{Ambrosetti2010a} we obtain that more segregated systems entail
lower values of $\phi_{1c}$. This last observation is in agreement with the behavior seen in real segregated
composites if we reinterpret the percolation threshold values reported in the literature as our crossover
concentration $\phi_{1c}$.

As we shall see in the next section, the interpretation that the segregation basically leads to a shortening of the
tunneling distances can be established on a firmer ground by employing the critical path approximation to the
tunneling network. This analysis will also provides us with useful explicit formulas for the overall 
composite conductivity.

\section{Critical path approximation}
\label{critical}

\begin{figure}[t!]
\begin{center}
\includegraphics[scale=0.36,clip=true]{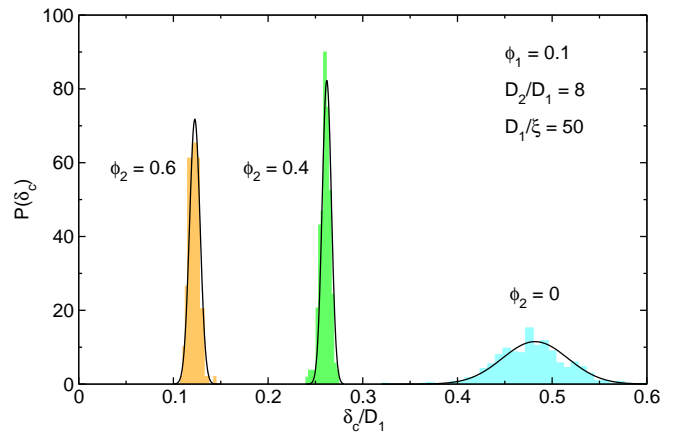}
\caption{(Color online) Histograms for the distribution $P(\delta$) of the minimum interparticle
distance $\delta$ required to have percolation for $500$ realizations of the same systems
considered in Fig.~\ref{fig2}.
The solid lines are fits to gaussian distribution functions.}\label{fig4}
\end{center}
\end{figure}

We show in this section that, as for homogeneous dispersions of spherical, rod-like, and plate-like 
impenetrable particles,\cite{Ambrosetti2009,Ambrosetti2010a} the critical path approximation (CPA) can 
also reproduce to a high accuracy the conductivity behavior for the inhomogeneous dispersions considered here.
According to the CPA,\cite{Ambegaokar1971} the composite conductivity $\sigma$ can be expressed approximately as
\begin{equation} \label{eq:cp}
 \sigma = \sigma_0 \exp \! \left( -\frac{2}{\xi} \delta_c \right),
\end{equation}
where $\delta_c \equiv \delta_c (\phi_1, \phi_2, D_2/D_1 )$ is the largest among the $\delta_{ij}$ distances such that the sub-network defined by all distances  $\delta_{ij}<\delta_c$ forms a conducting cluster spanning (or percolating) the entire sample.
The critical distance $\delta_c$ defines the single bond critical conductance $g_c =  \exp(-2\delta_c/\xi)$ which,
once assigned to all the conductances $g_{ij}$ of the network, leads to
Eq.~\eqref{eq:cp}, where the prefactor $\sigma_0$ is the only remaining fitting parameter.

Equation \eqref{eq:cp} reduces the conductivity problem to a simpler, geometrical one, which amounts to find
the geometrical critical distance $\delta_c$ so that percolation is established. In practice, $\delta_c$
can be obtained by coating each conducting sphere by a concentric penetrable shell of thickness $\delta/2$
and by considering two spheres as connected if their shells overlap. The critical distance $\delta_c$
is then the minimum value of $\delta$ such that (for given values of $\phi_1$, $\phi_2$, and $D_2/D_1$)
a cluster of connected spheres spans the sample.
Hence, in contrast to the usual core-shell model approach to transport,\cite{Wang1993,Jing2000,He2004,Berhan2007,Schilling2007,Kyrylyuk2008,Otten2009} here the shell 
thickness is not fixed \textit{a priori} but it depends on the particle concentration.

Our numerical procedure to find $\delta_c$ goes as follows. For fixed volume fraction $\phi_1$ of the conducting
spheres, as well as for given $\phi_2$ and $D_2/D_1$ values, we first generate the system as explained in
Sec.~\ref{model}. For each realization $i$ of the system ($i=1,\ldots,N_r$)
we chose an initial value $\delta_i$ comprised within the interval $\Delta\delta=\delta_{\rm max} - \delta_{\rm min}$, where
$\delta_{\rm min}=0$ and $\delta_{\rm max}$ is large enough so that a percolating cluster is surely established.
Clustering is performed on the adjacency list which represents the vicinity network.
Namely, we scan iteratively the list by checking if a given node already belongs to previously classified clusters.
If this condition is not fulfilled, we form a new cluster by identifying
which nodes are directly and indirectly connected to the selected node. This is done by labeling the
first and, recursively, the next levels in the vicinity hierarchy. Finally, a percolating cluster 
is the one in which at least two of its nodes lie at the opposite faces of the sample cube.
The critical distance $\delta_c^i$ for the $i$-th realization is then found by bisecting the interval $\Delta\delta$
until convergence is reached within a relative error of $10^{-3}$. Finally, an histogram representing the
distribution $P(\delta_c)$ of the critical distance is obtained by repeating the procedure for all the $N_r$
realizations of the system.

Examples of the thus obtained $P(\delta_c)$ are shown in Fig.~\ref{fig4} for the same cases of Fig.~\ref{fig2}
(i.e., $\phi_1=0.1$, $D_2/D_1=8$, and $\phi_2=0$, $0.4$, and $0.6$). They approximately follow normal
distributions (solid lines) centered at critical distances which steadily decrease as one moves
from the homogeneous case ($\phi_2=0$) to the increasingly segregated regimes ($\phi_2=0.4$ and $0.6$),
thus confirming our previous conjecture that segregation implies a shortening of the tunneling lengths.

\begin{figure}[t!]
\begin{center}
\includegraphics[scale=0.8,clip=true]{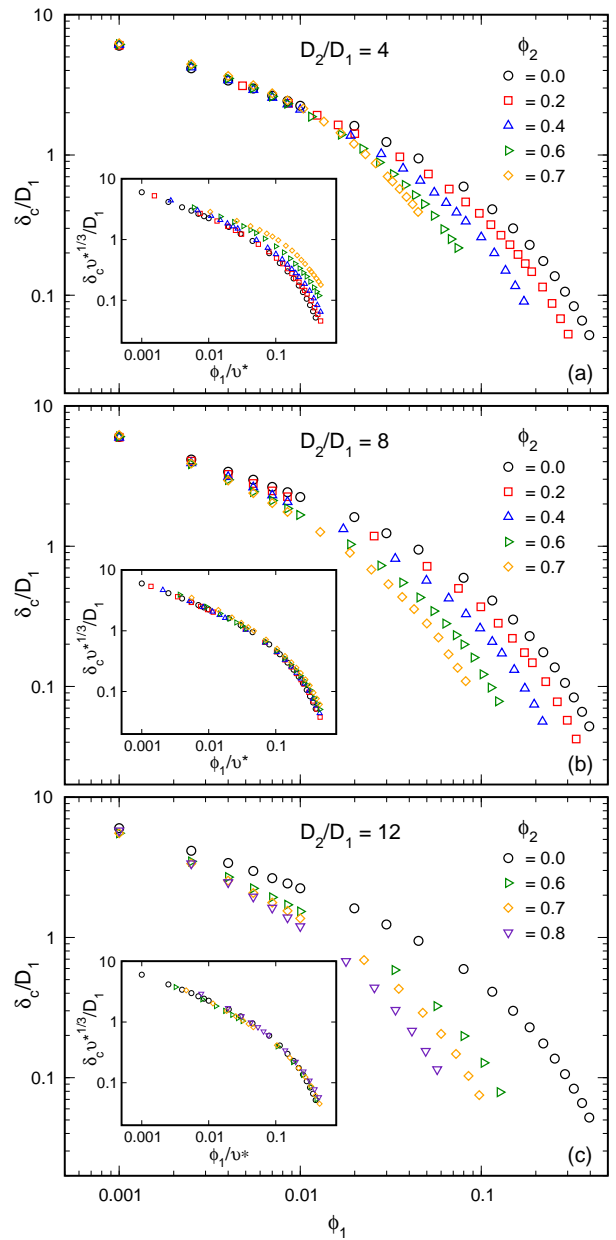}
\caption{(Color online) Critical distance $\delta_c$ dependence on the volume fraction $\phi_1$ of the
conducting $D_1$ spheres for several values $\phi_2$ of the volume fraction of the insulating
$D_2$ spheres for (a) $D_2/D_1=4$, (b) $D_2/D_1=8$, and (c) $D_2/D_1=12$. Insets: the same data
plotted according to the scaling relation of Eq.~\eqref{eq:scaling1}.}\label{fig5}
\end{center}
\end{figure}

This trend is clearly seen in Fig.~\ref{fig5}, where we plot the overall behavior of $\delta_c$ as a function
of $\phi_1$ and for different values of $\phi_2$ and $D_2/D_1$. At very low filler volume fractions, all
$\delta_c$ curves for $\phi_2\neq 0$ tend asymptotically to the critical distance of the homogeneous case
(open circles) which in this limit behaves as $\delta_c\propto \phi_1^{-1/3}$, indicating that segregation
is irrelevant for $\phi_1\rightarrow 0$. As $\phi_1$ increases, segregation acts by lowering the critical
distance which, for large $\phi_2$ and $D_2/D_1$ values, can be even one order of magnitude smaller than that
of the homogeneous case.

The reduction of $\delta_c$ can be explained by using the argument that the available volume for placing the
conducting spheres is reduced by segregation. Indeed, as Fig.~\ref{fig1}(b) suggests, such a reduction
has the net effect of increasing the local density of the $D_1$ particles, and so of reducing the critical
distance for percolation. This argument neglects conductor-insulator interface effects and assumes that
narrow bottlenecks are irrelevant for the establishment of the critical paths, which are conditions both fulfilled
in the $D_2/D_1\gg 1$ case.\cite{Johner2008}

\begin{figure*}[t!]
\begin{center}
\includegraphics[scale=0.7,clip=true]{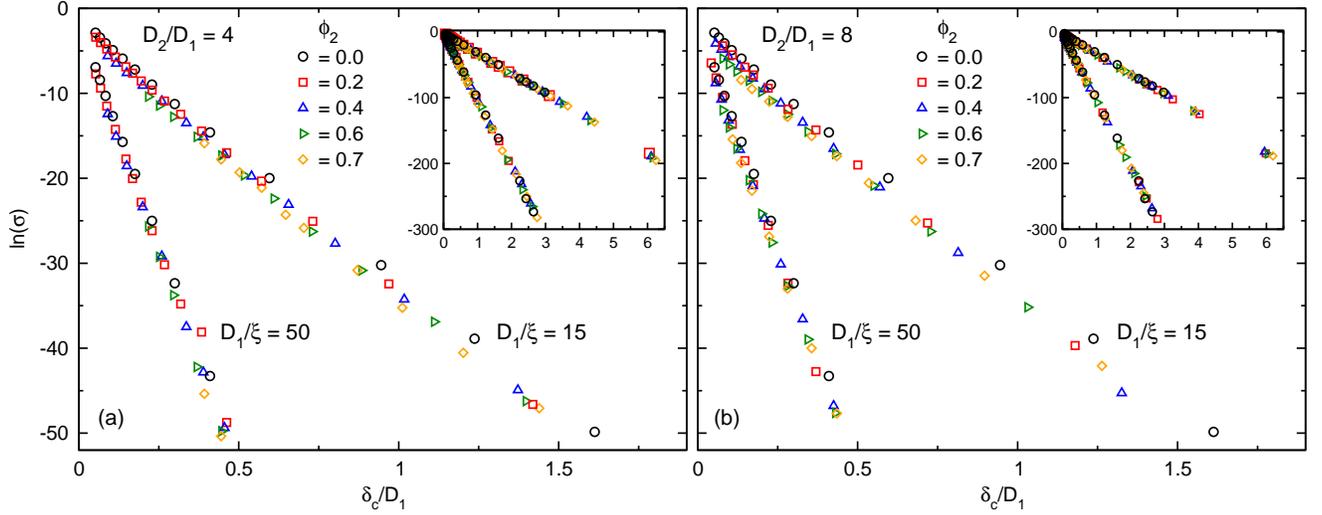}
\caption{(Color online) Natural logarithm of the conductivity values $\sigma$ of Fig.~\ref{fig3}
plotted as a function of the critical distance $\delta_c$ of Fig.~\ref{fig5}.
(a) $D_2/D_1=4$; (b) $D_2/D_1=8$. The insets shows the $\ln(\sigma)$ versus $\delta_c/D_1$ dependence in a
larger scale. }\label{fig6}
\end{center}
\end{figure*}

By following this line of reasoning, we argue that the only relevant variable for describing segregation is
the available volume fraction $\upsilon^*$, and that consequently the critical distance can be expressed
as $\delta_c(\phi_1,\upsilon^*)$. Furthermore, by requiring that the relevant $\phi_1$ dependence of the $D_1$
particles is through the effective volume fraction $\phi_1/\upsilon^*$, we express the critical distance
as $\delta_c(\phi_1,\upsilon^*)=a f(\phi_1/\upsilon^*)$, where $f(x)$ is a general function and $a$ is
a proportionality constant. The value of $a$ can be determined by requiring that the critical distance for $\phi_1\rightarrow 0$ coincides with that of the homogeneous case, which leads us to the following scaling relation
\begin{equation}
\label{eq:scaling1}
\delta_c(\phi_1,\upsilon^*)=\upsilon^{*-1/3}\,f(\phi_1/\upsilon^*).
\end{equation}
As shown in the insets of Fig.~\ref{fig5}, where $\upsilon^{*1/3}\delta_c$ is plotted as a function of
$\phi_1/\upsilon^*$, the data for different $\phi_2$ values basically collapse into a single curve already for
$D_2/D_1\ge 8$. However, for $D_2/D_1=4$ the scaling relation \eqref{eq:scaling1} is not very effective but, 
as shown in the Appendix, the scaling argument can be generalized in order to provide a better data collapse 
also for $D_2/D_1< 8$.

From the scaling relation \eqref{eq:scaling1} it follows that for $\upsilon^*=1$ the function $f(\phi_1)$
coincides with the critical distance in the homogeneous limit $\delta_c(\phi_1,1)$, thereby leading to
\begin{equation}
\label{eq:scaling}
\delta_c(\phi_1,\upsilon^*)=\upsilon^{*-1/3}\,\delta_c(\phi_1/\upsilon^*,1),
\end{equation}
which merely states that the critical distance in the segregated regime can be directly obtained from
that of the homogeneous case. In order to illustrate the consequences of Eq.~\eqref{eq:scaling} for the
conductivity $\sigma$ of segregated composites, let us first verify that the CPA of Eq.~\eqref{eq:cp} actually
provides a valuable approximation of $\sigma$. In Fig.~\ref{fig6} we replot the conductivity data of
Fig.~\ref{fig3} as a function of the critical distance results of Fig.~\ref{fig5}. Irrespectively of $\phi_2$
and of $D_2/D_1$, we find that $\sigma$ nicely follows the linear relation
\begin{equation}
\label{linear}
\ln(\sigma)=\ln(\sigma_0)-\frac{2}{\xi}\delta_c,
\end{equation}
with slope $2/\xi$, and that therefore the CPA is in excellent agreement with the full numerical
solution of the tunneling resistor network. Thus, from Eqs.~\eqref{eq:cp} and \eqref{eq:scaling}, this means 
that for $D_2/D_1\gg 1$ (see the Appendix for the $D_2/D_1\sim 1$ case) we
can express $\sigma$ in terms of the critical distance $\delta_c(\phi_1,1)$ of the homogeneous limit, leaving
only the prefactor $\sigma_0$ to be determined. By using some approximate formula for $\delta_c(\phi_1,1)$,
like for example those reported in Refs.~\onlinecite{Heyes2006,Ambrosetti2009,Ambrosetti2010a}, the full
$\phi_1$ dependence of $\sigma$ for given $\upsilon^*$ can be then expressed in analytical terms
with high accuracy. This is illustrated in Fig.~\ref{fig3}(b), where the solid lines have been obtained from
\begin{equation}
\label{eq:sigmafit}
\sigma=\sigma_0\exp\!\left[-\frac{2}{\xi}
\frac{\delta_c(\phi_1/\upsilon^*,1)}{\upsilon^{*1/3}}\right],
\end{equation}
with $\upsilon^*$ from Eq.~\eqref{eq:phiav}, $D_2/D_1=8$, and $\delta_c(x,1)$ as given in
Ref.~\onlinecite{Ambrosetti2009}.\cite{notedelta} As shown in the next section, the possibility of
expressing the composite conductivity in terms of an analytic formula is a valuable tool
to describe the filler dependences of real segregated composites and to interpreted the experimental data.

\section{Application to experiments}
\label{applic}
\begin{figure}[t!]
\begin{center}
\includegraphics[scale=0.3,clip=true]{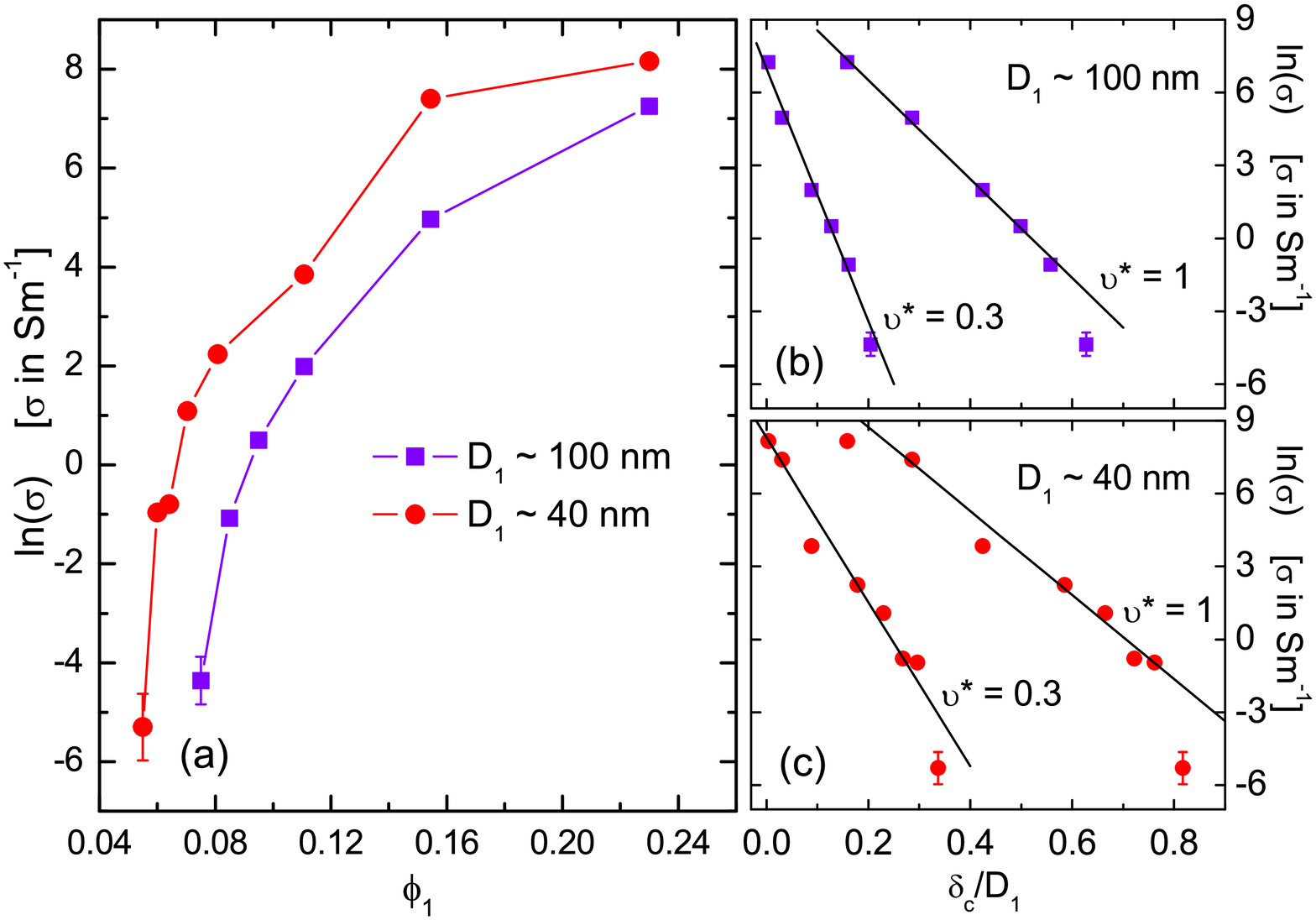}
\caption{(Color online) (a) Experimental conductivity $\sigma$ data of two series of RuO$_2$-cermets as a
function of the volume fraction $\phi_1$ of the RuO$_2$ particles (from Ref.~\onlinecite{Vionnet2005}).
The same $\sigma$ values are plotted as a function of the critical distance $\delta_c$ for
(a) $D_1\sim 100$ nm and (c) $D_1\sim 40$ nm by assuming two different values of the available
volume fraction $\upsilon^*$. The solid lines are fits to Eq.~\eqref{linear}.}\label{fig7}
\end{center}
\end{figure}

We show here that the result of the previous section can be used
to analyze experimental $\sigma$ vs $\phi_1$ data in order to extract estimates of the
tunneling length $\xi$ value and of the degree of segregation in the composite.
This is so because, if the experimental values of $\ln(\sigma)$ are plotted as a function of $\delta_c$
instead of $\phi_1$, and if our GTN picture applies, then they are expected to
follow Eq.~\eqref{linear} whose slope $-2/\xi$ directly gives the value of the tunneling
factor $\xi$ independently of the specific $\sigma_0$ value.
Furthermore, the available volume fraction $\upsilon^*$ appearing in Eq.~\eqref{eq:scaling}
does not depend, in principle, on the particular model chosen to represent segregation,\cite{noteupsilon1} 
and so it may used as a fitting parameter which best reproduce the experimental data.

In order to illustrate how the theory applies to real composites, we consider here the conductivity data
of RuO$_2$-cermet samples which were already reported in Ref.~\onlinecite{Vionnet2005}. In particular
we consider two series of samples constituted by RuO$_2$ conducting particles of mean sizes
$D_1\approx 40$ nm and $D_1\approx 100$ nm, dispersed in a borosilicate
glass. The glassy grains prior to thermal processing (firing) had average size of about $3\mu$m, so that 
for both series of composites $D_2/D_1\gg 1$. The two series of samples were fired by following identical 
thermal cycles so that, in principle, they differ only in the mean size $D_1$ of the conducting RuO$_2$ 
particles. It should be noted however that although the finer RuO$_2$ powders were given by nearly spherical 
and monodispersed particles, the coarser powders had more dispersed grain sizes with less regular shape.

In Fig.~\ref{fig7}(a) we plot the measured conductivity as a function of RuO$_2$ volume fraction $\phi_1$
for both series of composites. In Ref.~\onlinecite{Vionnet2005} we interpreted these same data in the
framework of percolation theory and fitted them with the power-law relation $(\phi_1-\phi_{1c})^t$.
The resulting low percolation threshold values, $\phi_{1c}\simeq 0.07-0.05$, were found to be consistent with the
segregated distribution of the RuO$_2$ conducting phase observed in the microstructure, while the large
transport exponent values, $t\simeq 3-4$, were concluded to arise from the nonuniversality of the critical
behavior as predicted by the tunneling-percolation model of Ref.~\onlinecite{Balberg1987} (see also later
developments of this theory in Refs.~\onlinecite{Grimaldi2006,Johner2008}).  Here, we offer an alternative
interpretation of these data based on the GTN theory which, as explained in this paper and in
Refs.~\onlinecite{Ambrosetti2009,Ambrosetti2010a,Ambrosetti2010b}, is more justified on physical grounds than our
previous percolation-based one of Ref~\onlinecite{Vionnet2005}.

\begin{figure}[t!]
\begin{center}
\includegraphics[scale=0.3,clip=true]{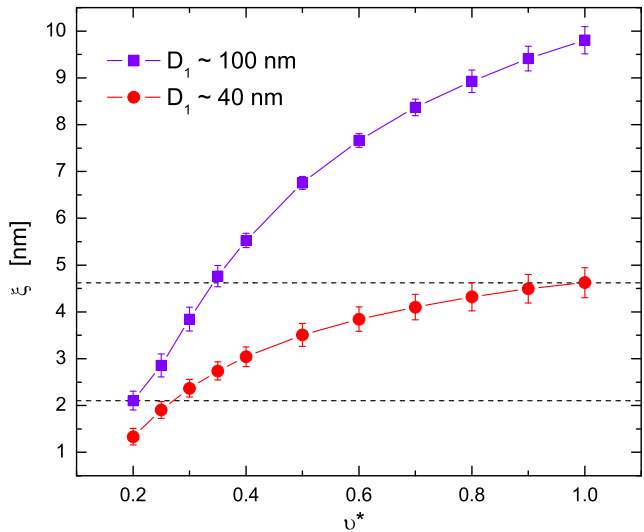}
\caption{(Color online) Values of the tunneling decay distance $\xi$ as a function of the available volume
fraction $\upsilon^*$ extracted as explained in the text. The two horizontal dashed lines at $4.6$ nm
and $2.1$ nm represent the maximum and minimum values of $\xi$ which are compatible with both series 
of composites.}\label{fig8}
\end{center}
\end{figure}

Let us start by re-plotting the conductivity data
of Fig.~\ref{fig7}(a) in terms of the critical distance $\delta_c(\phi_1,\upsilon^*)$ by using the scaling
relation of Eq.~\eqref{eq:scaling}.
For the functional form of $\delta_c(\phi_1/\upsilon^*,1)$ we use the fitting formula published in
Ref.~\onlinecite{Ambrosetti2009},\cite{notedelta} and we treat $\upsilon^*$ as an adjustable
parameter.\cite{noteupsilon2}
The resulting $\ln(\sigma)$ vs $\delta_c$ plots for two different values of $\upsilon^*$ are shown in
Fig.~\ref{fig7}(b) and Fig.~\ref{fig7}(c) for the $D_1\approx 100$ nm and $D_1\approx 40$ nm samples, respectively.
In both cases the data follow a better linear dependence for $\upsilon^*=0.3$ (strong segregation) than for $\upsilon^*=1$ (homogeneity), and from the corresponding fits to Eq.~\eqref{linear} (solid lines) we obtain that
lower values of $\upsilon^*$ imply also lower values of $\xi$ (i.e., the slopes are larger).
This is most clearly seen in Fig.~\ref{fig8} where we report the $\upsilon^*$
dependence of the so-obtained $\xi$ values for both series of samples. Starting from the homogeneous limit
at $\upsilon^*=1$ the tunneling length $\xi$ of the $D_1\approx 100$ nm series decreases by a factor of
five when the available volume fraction is lowered down to the minimum value $\upsilon^*=0.2$ for which
a linear fit of $\ln(\sigma)$ vs $\delta_c$ was possible, while the $D_1\approx 40$ nm
case displays a weaker decrease in the same range of $\upsilon^*$ due to the smaller RuO$_2$ grain size.
By realizing that the tunneling decay length should be independent of the size $D_1$ of the conducting particles,
while the specific $\upsilon^*$ value could be different for the two series of composites,
then $\xi$ must be comprised between the two horizontal dashed lines at $4.6$ nm and $2.1$ nm in Fig.~\ref{fig8}.
However, since the microstructure of both series of composites displays a marked segregated dispersion,\cite{Vionnet2005} then $\upsilon^*$ should be sensibly smaller than the unity,
suggesting that the lower limit $\xi\simeq 2$ nm is a more reliable estimate for $\xi$. 
This value is fully comparable to those extracted from other conductor-insulator
composites,\cite{Ambrosetti2009,Ambrosetti2010a} and, specifically, agrees well with the results of 
microscopical investigations of thick-film cermet resistors.\cite{Chiang1994}

\section{Conclusions}
\label{concl}

In this paper we have generalized the GTN model, where each conducting particle is connected to all
others through tunneling processes, to describe composites whose microstructure is
given by a segregated dispersion of the conducting phase. This particular class of nonhomogeneous
composites is characterized by large conductivity even for volume fraction values $\phi_1$ of the conducting
phase as low as a few percents. According to the percolation theory, this behavior is explained by the reduced
values of the percolation threshold induced by the segregated dispersion of the conducting particles. Here,
we have shown that the $\phi_1$ dependence of the conductivity in nonhomogeneous segregated composites can 
be understood without imposing any fixed cut-off in the microscopic electrical connectivity 
(as it is done in percolation theory) and that the GTN formulation provides thus a natural and physically justified 
approach to the study of transport in disordered composites.

Besides the full numerical solutions of the tunneling resistor network, we have also
shown that the critical path approximation is valid for a wide range of $\phi_1$, and that it permits to
formulate a scaling relation connecting the critical tunneling distance $\delta_c$ for a segregated systems
with that of a homogeneous composite. Finally, we have illustrated the practical importance of this scaling by
applying it to experimental conductivity data of RuO$_2$-cermet segregated composites, which has permitted
us to extract a realistic tunneling decay length $\xi$ and to estimate the degree of segregation in these materials.

\acknowledgements
This work was supported by the Swiss National Science Foundation (Grant No. 200021-121740).

\appendix
\section{improved scaling formula}

\begin{figure}[t]
\begin{center}
\includegraphics[scale=0.34,clip=true]{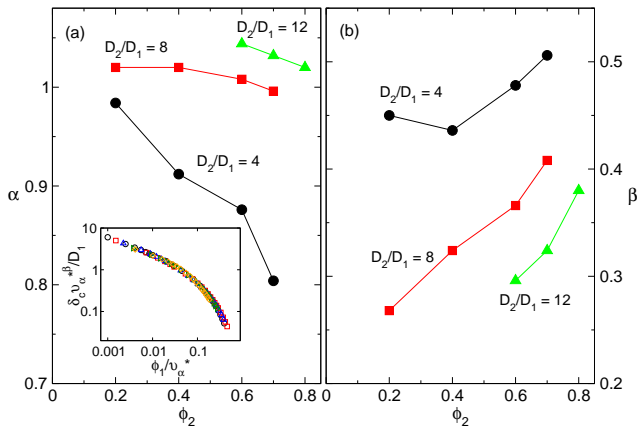}
\caption{(Color online) Dependence of the correction parameter (a) $\alpha$
and (b) $\beta$ on the volume fraction $\phi_2$ of the insulating spheres and on $D_2/D_1$.
Inset: the $\delta_c$ data for $D_2/D_1=4$ of Fig.~\ref{fig5}(b) rescaled according to
Eq.~\eqref{eq:scaling2} for the corresponding
values of $\alpha$ and $\beta$.}\label{fig9}
\end{center}
\end{figure}

Here we briefly address the problem of generalizing the scaling relation of Eq.~\eqref{eq:scaling} in order to extend
its validity beyond the $D_2/D_1\gg 1$ regime. To this end, let us first remind that $\upsilon^*$ defined in Eq.~\eqref{eq:phiav} gives the volume fraction available for placing the centers of the conducting particles with
diameter $D_1$. However, in order to define the critical distance $\delta_c$, the conducting spheres are
treated as core-shell particles, where the hard-core of diameter $D_1$ is coated by a concentric penetrable shell
of thickness $\delta_c/2$. Since this penetrable shell may actually overlap the insulating spheres then,
to what concerns the connectivity, these latter may be treated as having effectively a smaller diameter
$\alpha D_2\le D_2$, where $0\le\alpha\le 1$ is a correction factor which captures such effective reduction.
In this way the corrected available volume fraction reads:
\begin{equation}
\label{eq:phiav2}
\upsilon^*_{\alpha} = (1-\phi_2)^{(\alpha+D_1/D_2)^3}.
\end{equation}
For $\alpha\le 1$ the net effect of the overlapping between shells and insulating spheres is then an
effective increase of the available volume. This increase is expected to be unimportant when $D_2/D1 \gg 1$, so
that $\alpha\simeq 1$ in this regime, while when the two diameters $D_1$ and $D_2$ are comparable
the $\alpha$ correction has to be considered. By using Eq.~\eqref{eq:phiav2}, we generalize thus the scaling
relation \eqref{eq:scaling1} as follows:
\begin{equation}
\label{eq:scaling2}
\delta_c(\phi_1,\upsilon^*_\alpha)=\upsilon_\alpha^{*-\beta}\,f(\phi_1/\upsilon^*_\alpha),
\end{equation}
where in addition to the corrected available volume $\upsilon^*_{\alpha}$ we have introduced the new exponent
$\beta$ to improve the scaling in the large $\phi_1$ region.

In Fig.~\ref{fig9} we plot (a) $\alpha$ and (b) $\beta$ obtained from the
minimization of $|| \delta_c(\phi_1/\upsilon^*_{\alpha},1)- \upsilon_{\alpha}^{*-\beta} \delta_c||$ as
functions of the volume fraction $\phi_2$ of the insulating spheres and for $D_2/D_1 =4$, $8$, and $12$.
For $D_2/D_1=8$ and $12$ the coefficients $\alpha$ and $\beta$ are close to respectively $\alpha=1$
and $\beta=1/3$ for all values of $\phi_2$ considered, so that, as expected, Eq.~\eqref{eq:scaling} 
with $\upsilon^*$ as given in Eq.~\eqref{eq:phiav} provides a rather good scaling of the $\delta_c$ data. 
On the contrary, for $D_2/D_1=4$ the coefficient $\alpha$ displays a stronger $\phi_2$ dependence
and is sensibly smaller than $\alpha=1$ for large $\phi_2$ values, while $\beta$ is larger than $1/3$.
In particular, $\alpha< 1$  indicates that for this case the increase of the effective available 
volume $\upsilon^*_{\alpha}$ is an important effect for the correct scaling of $\delta_c$ which, as shown 
in the inset of Fig.~\ref{fig9}(a), is now almost perfect.

%In particular we affirmed that eq 4 does not holds when $\phi_2 >> 0$  and $D_2/D_1 \sim 1$. The former condition %refers to the structural formations of narrow necks in the volume [CITE] left over by the overlapping spheres [CITE %Feng , Halperin, Seng 1986] which makes $\delta_c$ dependent on both the connectivity of the necks and on their width. %The latter in some way is a measure of how the insulating dispersion can be seen as a continuum ($D_2/D_1>>0$) or a %granular ($D_2 \sim D_1 $) phase. The no-linear $\delta_c(\phi_2)$ effect becomes even more evident if $D_2/D_2 \sim 1 %$ [CITE NIKI]

%In this paper we did not approach the regime of high segregation and the non linear decrease of $\delta_c (\phi_2)$
%is evident only in fig NUM for $D_2/D_1 = 4$ and $\phi_2 = 0.7$.  We considered instead the case of comparable sphere's %diameters ($D_2/D_1=4$) when the scaled $\delta_c(\phi_2, D_2/D_1)$ curves poorly fit the homogeneous one [fig num] and %in general a correction in eq NUM is needed also for $D_2/D_1=8, 12$. The problem can be addressed through simple  %considerations about the system structure. \\


\begin{thebibliography}{38}
\bibitem{deltastar}
Specifically, we have first evaluated the critical bond conductance $g_c=\exp(-2\delta_c/\xi)$ resulting
from the critical path approximation described in Sec. \ref{critical}. Next, we have chosen $\delta^*$ in order to
satisfy $\exp(-2\delta^*/\xi)=\exp(-M)g_c$, where  $M=20$, $10$, $5$
for $g_c<10^{-50}$, $10^{-50}<g_c<10^{-10}$, and $10^{-10}<g_c$, respectively.
\bibitem{notedelta}
Note that in the expression for $\delta_c(x,1)$ given in Ref.~\onlinecite{Ambrosetti2009}
there is a factor $2$ missing. The correct expression is: $\delta_c(x,1)=
2/[A_0+A_1\exp(x/t_1)+A_2\exp(x/t_2)]$, with $A_0=-7.33$, $A_1=7.25$, $A_2=0.81$,
$t_1=0.355$, and $t_2=0.114$.
\bibitem{noteupsilon1}
Indeed, the concept of available volume fraction is very general, and it can be applied also
to dispersions of insulating objects which are non-spherical, polydispersed, partially or completely
impenetrable.
\bibitem{noteupsilon2}
Since the larger RuO$_2$ volume fraction value in Fig.~\ref{fig7}(a) is $\phi_1=0.23$, for $\upsilon^*<0.36$ 
the effective concentration
$\phi_1/\upsilon^*$ becomes larger than the random close-packed limit $\simeq 0.64$.
Hence, in order to maintain consistency with our formulation, we have considered in the fitting for $\upsilon^*<0.4$
only the low density data of Fig.~\ref{fig7}(a).
\end{thebibliography}
\end{document}